\documentstyle[11pt,IAU207_pasp,twoside]{article}
\def\edcomment#1{\iffalse\marginpar{\raggedright\sl#1\/}\else\relax\fi}
\reversemarginpar

\begin{document}
\title{The determination of the age of Globular Clusters:\\
a statistical approach}
 \author{Miriam Rengel $^{1,2}$}
\affil{$^{1}$TLS Tautenburg, 07778 Tautenburg, Germany
}
\affil{$^2$Centro de Investigaciones de Astronom\'{\i}a, M\'erida 5101-A, Venezuela}
\author{Juan Mateu}
\affil{Centro de Investigaciones de Astronom\'{\i}a, M\'erida 5101-A, Venezuela}

\author{Gustavo Bruzual}
\affil{Centro de Investigaciones de Astronom\'{\i}a, M\'erida 5101-A, Venezuela}

\begin{abstract}

\noindent
We present a statistical approach for determining the age of Globular
Clusters (GCs) that allows estimating the age derived from CMDs more
accurately than the conventional methods of the isochrone fitting. We
measure how closely a set of synthetic CMDs constructed from different
evolutionary models resemble the observed ones by determining the
likelihood using Saha and $\chi^2$ statistics. The model which best
matches the observational data, of a set of plausible ones, is the one
with the highest value of the estimator. We apply this method to a set
of three different evolutionary models presented by three different
authors. Each of theses sets consists again of many different models of
various chemical abundances, ages, input physics and there alike. We
subsequently derive the age of GCs NGC 6397, M92 and M3. With a
confidence level of 99\%, we find that the best estimate of the age is
14.0 Gyrs within the range of 13.8 to 14.4 Gyrs for NGC 6397, 14.75 Gyrs
within the range of 14.50 to 15.40 Gyrs for M92, and 16.0 Gyrs within
the range of 15.9 to 16.3 Gyrs.

\end{abstract}

\section{Introduction}

The age of Globular Clusters (GCs) has been recognized as being of key
importance for deriving the age of the Universe. However, up to now,
determining the reliable age of GCs has been proven rather difficult. This
is not only because of errors introduced by the observations, but also
because of rather subjective way in which the stellar evolutionary
models have been selected in some conventional procedures like the
isochrone fitting and the way in which observational data has been
fit. In this work, we present a statistical approach for determining
of the age of GCs that overcomes some of theses difficulties. Our
method avoids the difficulty in selecting the stellar evolutionary model
which best matches the observational data with the stellar model of the
CMD. This is done by determining the likelihood using the Saha statistics
or {\it W} (Saha 1998). The goodness of the fit of our method is estimated from
the $\chi^2$ statistics by Press et al. 1986. As an example, we apply
the method to NGC 6397, M92 and M3, and give an estimate for the
age of them with a confidence level of 99\%.

\begin{figure}
\begin{center}
\vskip6.8truecm
\includegraphics{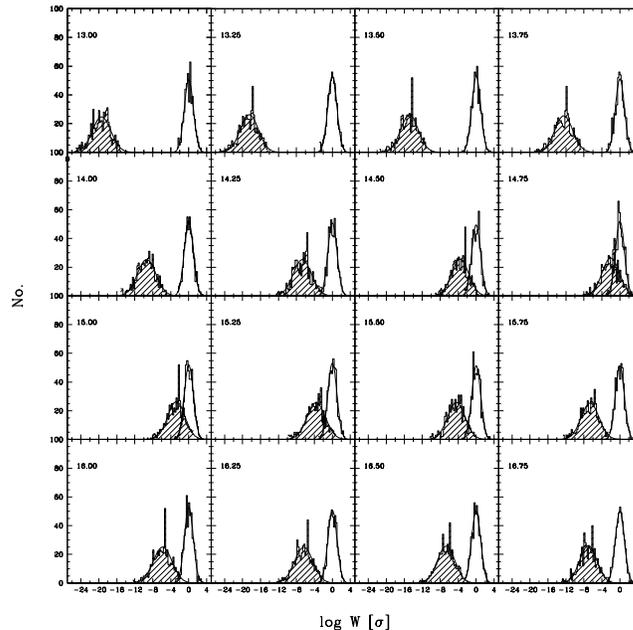}
\end{center}
\caption[]{Distribution of values of log {\it W} for different
ages of the Padova isochrones for stars in NGC 6397
(see text for details).}
\label{w-values}
\end{figure}

\begin{table}[]
\begin{center}
\caption{Summary of the results of the best estimation of the age of the
samples
considered in this work, for every evolutionary model}
{\scriptsize
\begin{tabular}{cccccc}
 
Evolutionary & Z of the & $\chi^{2}_{min}$~~~~ & $\chi^{2}_{red}$ & Best estimation&Interval of 99\%\\
Model  & Isochrone  &  &  & of t [Gyrs] &of confidence\\
\noalign{\smallskip}
\hline
\noalign{\smallskip}
NGC 6397        &1187 stars      &B=14400&      &         &
       \\[0mm]
Padua           &0.0004          &1352.81& 2.82 &14.50    & [14.3 - 15.1]     \\[0mm]
Yale            &0.0004          &1089.04& 2.34 &14.00    & [13.8 - 14.3]     \\[0mm]
\hline
NGC 6397        &373 stars      &B=8400  &      &         &
       \\[0mm]
Padua           &0.0004          &232.84 & 1.41 &14.25    & [13.7 - 15.6]     \\[0mm]
Yale            &0.0004          &206.80 & 1.25 &14.00    & [13.3 - 15.1]     \\[0mm]
Pisa            &0.0002          &249,95 & 1.41 &13.00    & [11.9 - 14.2]     \\[0mm]
\hline
M92             &4846 stars     &B=12000 &      &         &
      \\[0mm]
Padua           &0.0001         &2221.63 & 1.31 &14.75    & [14.5 - 15.4]     \\[0mm]
Yale            &0.0002         &2455.26 & 1.42 &15.00    & [14.6 - 15.7]     \\[0mm]
\hline
M92             &1482 stars     &B=12000  &      &         &
       \\[0mm]                                                                    
Padua           &0.0001         &678.79  & 1.73 &15.00    & [14.8 - 15.5]     \\[0mm]
Yale            &0.0002         &769.67  & 1.91 &15.00    & [14.7 - 15.5]     \\[0mm]
Pisa            &0.0002         &832.86  & 2.06 &12.00    & [11.8 - 12.1]     \\[0mm]
\hline
M3              &10333 stars    &B=14400  &      &         &
       \\[0mm]
Padua           &0.0004         &3558.38 & 1.59 &15.75    & [15.7 - 15.9]     \\[0mm]
Yale            &0.0004         &3037.01 & 1.37 &16.00    & [15.9 - 16.3]     \\[0mm]
\hline
M3              &4929 stars     &B=14400  &      &         &
       \\[0mm]
Padua           &0.0004         &1038.53 & 2.14 &16.25    &  [16.2 - 16.6]     \\[0mm]
Yale            &0.0004         &757.20  & 1.68 &16.00    &  [15.9 - 16.3]     \\[0mm]
Pisa            &0.0002         &1120.65 & 2.45 &14.00    &  [13.9 - 14.2]     \\[0mm]
\noalign{\smallskip}
\hline                                                                            
\noalign{\smallskip}
\end{tabular}}
\end{center}
\end{table}                 

\section{Observational data and stellar evolutionary models}

The observational data consist of V and I photometry of three GCs of the
Galaxy: NGC 6397 (D'Antona 1999 \& King 1999), M92 and M3 (JKT data by
Rosenberg 1999). We removed all stars with excessive photometric errors
(3 $\sigma$ in the colour) and we selected only stars out to a radius of
r$<$140$'$ for M92 and r$<$170$'$ for M3. The number of stars in each
sample is given in Tab.~1. We selected the models computed by Bertelli
et al. 1994 and Girardi et al 1996 (Padova isochrones), tracks developed
by Demarque et al (Yale isochrones) and a set of models by Cassini et
al.  (1998, 1999; Pisa isochrones). We chose them as they were published
in the theoretical plane and we use the transformations computed by Bruzual
\& Charlot 2000 for convert them to the observational plane.

\section{Results}

The full details of the approach are presented in Rengel 2000.
As an example, 
Fig. 1 shows the distribution of values of log {\it W} for
a sample of 1187 stars in NGC 6397. The non-shaded regions represent the
distribution of frequencies of log {\it W} obtained from 500 model-model
comparisons, the shaded regions for 500 data-model comparisons. The
estimation of the age is given for the age of the isochrone that
corresponds to the minimal distance obtained between the Gaussian median fits to
the model-model and data-model distributions (14.75 Gyrs). A summary of
the results obtained for the best estimation of the age of the GCs
for each of the models is given in Tab.~1.

\acknowledgments

We thank to A. Rosenberg, I. King and F. D'Antona for providing us the
data \& to G. Magris and X. Hernandez for
the discussions.

\end{document}